# Three-component fermions with surface "Fermi arcs" in topological semimetal tungsten carbide

J.-Z. Ma,[1,2,†] J.-B. He,[1,†] Y.-F. Xu,[1,2,†] B.-Q. Lv,[1,2] D. Chen,[1,2] W.-L. Zhu,[1,2] S. Zhang,[1] L.-Y. Kong,[1,2] X. Gao,[1,2] L.-Y. Rong,[2,3] Y.-B. Huang,[3] P. Richard,[1,2,4] C.-Y. Xi,[5] Y. Shao,[1,2] Y.-L. Wang,[1,2,4] H.-J. Gao,[1,2,4] X. Dai,[1,2,4] C. Fang,[1] H.-M. Weng,[1,4] G.-F. Chen,[1,2,4,*] T. Qian,[1,4,*] and H. Ding[1,2,4,*]

[1] *Beijing National Laboratory for Condensed Matter Physics and Institute of Physics, Chinese Academy of Sciences, Beijing 100190, China*

[2] *School of Physics, University of Chinese Academy of Sciences, Beijing 100190, China*

[3] *Shanghai Synchrotron Radiation Facility, Shanghai Institute of Applied Physics, Chinese Academy of Sciences, Shanghai 201204, China*

[4] *Collaborative Innovation Center of Quantum Matter, Beijing, China*

[5] *Anhui Province Key Laboratory of Condensed Matter Physics at Extreme Conditions, High Magnetic Field Laboratory of the Chinese Academy of Sciences, Hefei 230031, Anhui, China*

[†] These authors contributed to this work equally.

[*] Corresponding authors: dingh@iphy.ac.cn, tqian@iphy.ac.cn, gfchen@iphy.ac.cn

# Abstract

Topological Dirac and Weyl semimetals not only host quasiparticles analogous to the elementary fermionic particles in high-energy physics, but also have nontrivial band topology manifested by exotic Fermi arcs on the surface. Recent advances suggest new types of topological semimetals, in which spatial symmetries protect gapless electronic excitations without high-energy analogy. Here we observe triply-degenerate nodal points (TPs) near the Fermi level of WC, in which the low-energy quasiparticles are described as three-component fermions distinct from Dirac and Weyl fermions. We further observe the surface states whose constant energy contours are pairs of "Fermi arcs" connecting the surface projection of the TPs, proving the nontrivial topology of the newly identified semimetal state.

**One Sentence Summary**: Photoemission proves the nontrivial topology of a new type of semimetal that hosts three-component fermions .

The Standard Model predicts three types of fermionic elementary particles in space, *i.e.* Dirac, Weyl, and Majorana fermions. So far, only Dirac fermions have been identified and the existence of Weyl and Majorana fermions are still debated. In recent years, it has been realized that crystal symmetry-protected gapless electronic excitations in condensed matter systems are analogous to these elementary particles (*1-4*). For example, the excitations near 4- and 2-fold degenerate nodal points in the electronic structures can be described by the Dirac and Weyl equations, respectively (*4,5*). Materials hosting Dirac and Weyl nodes near the Fermi energy, called Dirac and Weyl semimetals, have been predicted theoretically and discovered experimentally (*5-12*). They manifest a nontrivial topology with surface Fermi arcs connecting to the projection points of bulk Dirac or Weyl nodes. These discoveries in condensed matter systems not only provide a platform for studying fundamental physics and emergent quantum phenomena of Dirac and Weyl fermions, but may also lead to practical applications in electronic devices.

In addition to the 4- and 2-fold degeneracy, it has been shown that spatial symmetries in crystals may protect other types of band degeneracy. The quasiparticle excitations near these degenerate points are not the analogues of any elementary fermions described in the Standard Model. Theory has proposed 3-, 6-, and 8-fold degeneracy protected by nonsymmorphic symmetries (*13*), and 3-fold degeneracy protected by symmorphic rotation plus mirror symmetries (*14-18*). Among them, the 3-fold degeneracy lies in an intermediate state between Dirac and Weyl states, as illustrated in Fig. 1A. Tuning protected symmetries induces topological phase transitions, which may lead to novel quantum phenomena. For example, breaking inversion symmetry in a Dirac semimetal may split the Dirac node into a pair of triply-degenerate nodal points (TPs), while breaking time-reversal symmetry in a proper way drives a TP into two Weyl nodes (*16,18*). Distinct from the isolated Dirac and Weyl nodes, the pair of TPs in Fig. 1A are connected by a doubly-degenerate nodal line (*14,17*), inducing touching Fermi surface (FS) pockets containing TPs. The topological invariant of individual pockets is not well defined, leading to the non-straightforward topological nature of the TP semimetal, in contrast to topological Dirac and Weyl semimetals.

First-principles calculations predict that the TPs exist in the electronic structure of a series of materials with the WC-type crystal structure (*15-18*). The presence of TPs was firstly demonstrated by our previous angle-resolved photoemission spectroscopy (ARPES) measurements on MoP (*19*), which is isostructural to WC. MoP has been recently reported to have extremely low resistivity and high carrier mobility, though the TPs are far below the Fermi level ($E_F$) and the three-component fermions are thought to have little contribution to the low-energy quasiparticle excitations (*20*). Since the topological surface states (SSs) are not observed in MoP (*19*), the evidence of the topological nature for the semimetal state is still lacking. In this work, we use ARPES to investigate the electronic structure of WC, and clearly observe three-component fermions near the Fermi energy, which would contribute to transport properties. More significantly, we identify robust surface Fermi arcs associated with the TPs, proving the nontrivial topology of the semimetal state in WC.

Our band structure calculations of WC indicate four TPs distributed on the $\Gamma$-A line (Fig. 1C). Each TP is formed by the crossing of one doubly-degenerate band and one non-degenerate band along $\Gamma$-A due to their band inversion (Fig. 1, C and F). In the plane perpendicular to $\Gamma$-A, three non-degenerate bands converge at the TP (Fig. 1, G and H). Furthermore, the strong spin-orbit coupling of W 5*d* electrons leads to large band splitting, which facilitates the identification of TPs in the ARPES measurement and the investigation of the associated topological nature. These advantages indicate that WC is an ideal platform to study the new fermions.

The measured samples have a typical triangular shape. The triangle surface is the

(001) plane and its edge is along the [010] direction, as illustrated in the inset of Fig. 1D. We obtained mirror-like (001) and (100) surfaces in a macroscopic scale (~ 0.1 × 0.1 mm$^2$) by cleavage. However, the ARPES measurements with vacuum ultraviolet (VUV) light did not detect any information of band structures on both surfaces. We did not observe clear band dispersions with photon energy smaller than 300 eV. To understand this phenomenon, we measured the (100) cleavage surface with scanning tunneling microscopy (STM). The ordered arrangement of atoms is not observed at the top layer (Fig. 4G), though the cleavage surfaces appear to be specular. While the photoelectrons excited by VUV light have a mean free path of a few Å, those excited by soft x-ray have a much longer escape depth (*21*). Therefore, the band structure detected in the soft x-ray ARPES experiments should reflect the electronic states beneath the disordered top layer.

The measured electronic structure exhibits a periodic modulation upon varying the photon energy of soft x-ray. Figure 2, D and E, plots the FSs recorded with two different photon energies on the (001) surface. They are consistent with the calculated bulk FSs in the $k_z = 0$ (Fig. 2H) and $\pi$ (Fig. 2I) planes, respectively. Moreover, we have performed quantum oscillation measurements with magnetic fields up to 38 Tesla (Fig. 2C). The extracted extreme cross sections match well with the four FS pockets at $k_z = 0$ and $\pi$ (Fig. 2, D and E). Similarly, the FSs measured on the (100) surface (Fig. 2, E and F) are consistent with the calculated bulk FSs in the $k_y$-$k_z$ plane (Fig. 2I). In addition, we observe almost straight FS lines connecting to the FSs near A on the (100) surface (Fig. 2E). The FS lines are absent in the calculated bulk FSs, whereas they are reproduced as the SSs in our slab calculations (Fig. 4B). Next we demonstrate the presence of TPs in the bulk electronic structure of WC (Fig. 3) and discuss the topological nature of the SSs (Fig. 4).

The band calculations show that only TP #1 is located below $E_F$ (Fig. 1C). A schematic three-dimensional (3D) plot of the band structure near TP #1 in the $k_x = 0$ plane is displayed in Fig. 3A. To search for the theoretically proposed TP #1, we have systematically measured the band dispersions on the (100) surface. We identify three electron bands along Γ-A (Fig. 3, C and D), which are consistent with the calculations (Fig. 3D). According to the Kramers theorem, the bands must be spin-degenerate at the time-reversal invariant momentum A. The two deep bands are degenerate at A and exhibit Rashba-like splitting upon dispersing towards Γ. In contrast, the shallow band remains doubly degenerate along the Γ-A high-symmetry line, which is protected by the vertical mirror symmetry. We clearly observe the band crossing of the doubly-degenerate shallow band and non-degenerate inner deep band, demonstrating the presence of TP #1 in the electronic structure of WC.

To further characterize TP #1, we measured the band dispersions perpendicular to Γ-A. Figure 3, F to J, shows five representative cuts at different $k_z$ values, whose momentum locations are illustrated in Fig. 3, A and B. We observe that the electron and hole bands touch at 200 meV below $E_F$ along cut 3 (Fig. 3, H and E). Furthermore, the results in Fig. 3, H and E exhibit two electron bands along cut 3, demonstrating the presence of TP #1 again.

We further prove the topological nature of the semimetal state in WC by identifying the surface Fermi arcs associated with the TPs on the (100) surface. As mentioned above, the observed FS lines are attributed to the (100) SSs. Since the STM topography image of the *in-situ* cleaved (100) surface shows no ordered arrangement of atoms at the top layer (Fig. 4G), the SSs should lie at the interface beneath the disordered top layer. The observed SSs appear to be very robust since they are not destroyed by the coverage of the disordered top layer.

The intensity map at 200 meV below $E_F$, which is at the energy of TP #1, exhibits the FS lines entering into bulk states around the projection of TP #1 (Fig. 4A). This implies that the SSs are associated with the TPs. FS lines also exist as the SSs in our slab calculations (Fig. 4B), though the calculations cannot simulate the real surface covered by a disordered layer. The calculations indicate that the FS lines are two pairs of Fermi arcs, which arises from two nearly parallel SS bands, as shown in Fig. 4E. Our experimental data also show a slight band splitting (Fig. 4, C and D), indicating that the observed FS lines consist of two pairs of Fermi arcs. The two pairs of Fermi arcs are consistent with the zero $Z_2$ number in the redefined "$k_z = 0$" plane (see Fig. S2 in the supplementary materials).

In contrast, the $k_z = \pi$ plane of WC has a nonzero $Z_2$ number (Fig. S2 in the supplementary materials). There must be an odd number of surface Dirac bands in the $k_z = \pi$ plane. We observe one hole band just below the bulk electron bands at $\tilde{A}$ (Fig. 4H). The hole band is also absent in the calculated bulk band structure, and thus belongs to the (100) SSs. Assuming that the hole band is the lower branch of the surface Dirac band, the upper branch should be very close to the conduction bands of bulk states (Fig. 4I), and thus is hard to be identified. However, the nontrivial topology of the $k_z = \pi$ plane guarantees the existence of the upper branch. The FS of the upper branch consists of either a closed circle enclosing $\tilde{A}$ or a pair of Fermi arcs connecting two TP #1 across the surface BZ boundary, as illustrated in Fig. 4J.

With the observation of surface Fermi arcs associated with the TPs in WC, we prove the topologically nontrivial nature of the new type of semimetal state beyond Dirac and Weyl semimetals. Our results suggest that three pairs of Fermi arcs emanate from the surface projection of one TP. It should be noted that while the SSs are topologically robust due to the nonzero $Z_2$ number of the $k_z = \pi$ plane, the Fermi arcs are not topologically protected and can evolve into closed FSs or even disappear, which is similar to the case in Dirac semimetals (*22-24*). However, our results show that the connection between the Fermi arcs and TPs in WC is not disturbed by the strong surface perturbation.

As the chemical potential of WC is located between a pair of TPs (TP #1 and #2), the three-component fermions must have significant contribution to the low-energy quasiparticles, which is favorable to the emergence of associated novel quantum phenomena. Very recently, we have observed quite strong anisotropic magnetoresistance in WC (*25*), which is likely to be related to the behavior of three-component fermions under magnetic fields applied along different directions, as theoretically proposed (*16*). This is in sharp contrast with the isotropic magnetoresistance in Dirac and Weyl semimetals (*26,27*). Our results suggest WC as an ideal platform for studying the topological quantum properties of three-component fermions.

**References:**

[1] A. H. Castro Neto, F. Guinea, N. M. R. Peres, K. S. Novoselov, A. K. Geim, The electronic properties of graphene. *Rev. Mod. Phys.* **81**, 109-162 (2009).

[2] L. Fu, C. L. Kane, Superconducting proximity effect and majorana fermions at the surface of a topological insulator. *Phys. Rev. Lett*. **100**, 096407 (2008).

[3] Z. Fang, N. Nagaosa, K. S. Takahashi, A. Asamitsu, R. Mathieu, T. Ogasawara, H. Yamada, M. Kawasaki, Y. Tokura, K. Terakura, The anomalous hall effect and magnetic monopoles in momentum space. *Science* **302**, 92 (2003).

[4] X. Wan, A. M. Turner, A. Vishwanath, S. Y. Savrasov, Topological semimetal and Fermi-arc surface states in the electronic structure of pyrochlore iridates. *Phys. Rev. B* **83**, 205101 (2011).

[5] Z. Wang, Y. Sun, X.-Q. Chen, C. Franchini, G. Xu, H.-M. Weng, X. Dai, Z. Fang, Dirac semimetal and topological phase transitions in $A_3$Bi (A = Na, K, Rb). *Phys. Rev. B* **85**, 195320 (2012).

[6] Z. K. Liu, B. Zhou, Y. Zhang, Z.-J. Wang, H. M. Weng, D. Prabhakaran, S.-K. Mo, Z.-X. Shen, Z. Fang, X. Dai, Z. Hussain, Y.-L. Chen, Discovery of a Three-Dimensional Topological Dirac Semimetal $Na_3Bi$. *Science.* **343**, 864-867 (2014).

[7] H. M. Weng, C. Fang, Z. Fang, B. A. Bernevig, X. Dai, Weyl Semimetal Phase in Noncentrosymmetric Transition-Metal Monophosphides. *Phys. Rev. X* **5**, 011029 (2015).

[8] S.-M. Huang, S.-Y. Xu, I. Belopolski, C.-C. Lee, G. Chang, B.-K. Wang, N. Alidoust, G. Bian, M. Neupane, C. Zhang, S. Jia, A. Bansil, H. Lin, M. Z. Hasan, A Weyl Fermion semimetal with surface Fermi arcs in the transition metal monopnictide TaAs class. *Nat. Commun* **6**, 7373 (2015).

[9] B. Q. Lv, H. M. Weng, B. B. Fu, X. P. Wang, H. Miao, J. Ma, P. Richard, X. C. Huang, L. X. Zhao, G. F. Chen, Z. Fang, X. Dai, T. Qian, H. Ding, Experimental Discovery of Weyl Semimetal TaAs. *Phys. Rev. X* **5**, 031013 (2015).

[10] S. Y. Xu, I. Belopolski, N. Alidoust, M. Neupane, G. Bian, C. Zhang, R. Sankar, G. Chang, Z. Yuan, C. C. Lee, S. M. Huang, H. Zheng, J. Ma, D. S. Sanchez, B. Wang, A. Bansil, F. Chou, P. P. Shibayev, H. Lin, S. Jia, M. Z. Hasan, Discovery of a Weyl fermion semimetal and topological Fermi arcs. *Science* **349**, 613 (2016).

[11] B. Q. Lv, N. Xu, H. M. Weng, J. Z. Ma, P. Richard, X. C. Huang, L. X. Zhao, G.F. Chen, C. E. Matt, F. Bisti, V. N.Strocov, J. Mesot, Z. Fang, X. Dai, T. Qian, M. Shi, H. Ding, Observation of Weyl nodes in TaAs. *Nat. Phys.* **11**, 724 (2015).

[12] L. X. Yang, Z. K. Liu, Y. Sun, H. Peng, H. F. Yang, T. Zhang, B. Zhou, Y. Zhang, Y. F. Guo, M. Rahn, D. Prabhakaran, Z. Hussain, S.-K. Mo, C. Felser, B. Yan, Y. L. Chen, Weyl semimetal phase in the non-centrosymmetric compound TaAs. *Nature Phys.* **11**, 728-732 (2015).

[13] B. Bradlyn, J. Cano, Z. Wang, M. G. Vergniory, C. Felser, R. J. Cava, B. A.

Bernevig, Beyond Dirac and Weyl fermions: Unconventional quasiparticles in conventional crystals. *Science* **353**, 558 (2016).

[14] T. T. Heikkilä, G. E. Volovik, Nexus and Dirac lines in topological materials. *New J. Phys.* **17**, 093019 (2015).

[15] H. Weng, C. Fang, Z. Fang, X. Dai, Coexistence of Weyl fermion and massless triply degenerate nodal points. *Phys. Rev. B* **94**, 165201 (2016).

[16] H. Weng, C. Fang, Z. Fang, X. Dai, Topological semimetals with triply degenerate nodal points in θ-phase tantalum nitride. *Phys. Rev.* B **93**, 241202 (2016).

[17] Z. Zhu, G. W. Winkler, Q. Wu, J. Li, A. A. Soluyanov, Triple point topological metals. *Phys. Rev. X* **6**, 031003 (2016).

[18] G. Chang, S.-Y. Xu, S.-M. Huang, D. S. Sanchez, C.-H. Hsu, G. Bian, Z.-M. Yu, I. Belopolski, N. Alidoust, H. Zheng, T.-R. Chang, H.-T. Jeng, S. A. Yang, T. Neupert, H. Lin, M. Z. Hasan, New fermions on the line in topological symmorphic metals. *arXiv*:1605.06831v1 (2016).

[19] B. Q. Lv, Z.-L. Feng, Q.-N. Xu, J.-Z. Ma, L.-Y. Kong, P. Richard, Y.-B. Huang, V. N. Strocov, C. Fang, H.-M. Weng, Y.-G. Shi, T. Qian, H. Ding, Experimental Observation of Three-Component 'New Fermions' in Topological Semimetal MoP. *arXiv*:1610.08877 (2016).

[20] C. Shekhar, Y. Sun, N. Kumar, M. Nicklas, K. Manna, V. Suess, O. Young, I. Leermakers, T. Foerster, M. Schmidt, L. Muechler, P. Werner, W. Schnelle, U. Zeitler, B. H. Yan, S. S.P. Parkin, C. Felser, Extremely high conductivity observed in the unconventional triple point fermion material MoP. *arXiv*:1703.03736 (2017).

[21] V. N. Strocov, M. Shi, M. Kobayashi, C. Monney, X. Wang, J. Krempasky, T. Schmitt, L. Patthey, H. Berger, P. Blaha, Three-dimensional electron realm in VSe$_2$

by soft X-ray photoelectron spectroscopy: origin of charge-density waves. *Phys. Rev. Lett*. **109**, 086401 (2012).

[22] S.-Y. Xu, C. Liu, S.-K. Kushwaha, R. Sankar, J. W. Krizan, I. Belopolski, M. Neupan, G. Bian, N. Alidoust, T.-R. Chang, H.-T. Jeng, C.-Y. Huang, W.-F. Tsai, H. Lin, P. P. Shibayev, F. Chou, R. J. Cava, M. Z. Hasan, Observation of Fermi arc surface states in a topological metal. *Science* **347**, 294 (2015).

[23] M. Kargarian, M. Randeria, Y.-M. Lu, Are the surface Fermi arcs in Dirac semimetals topologically protected? *Proc. Natl. Acad. Sci. U.S.A.*, **113**, 8648-8652 (2016).

[24] C. Fang, L. Lu, J. Liu, L. Fu, Topological semimetals with helicoid surface states. *Nat. Phys*. **12**, 936-941 (2016).

[25] J. B. He, D. Chen, W. L. Zhu, S. Zhang, L. X. Zhao, Z. A. Ren, G. F. Chen, Magnetotransport properties of the triply degenerate node topological semimetal: tungsten carbide. *arXiv*:1703.03211 (2017).

[26] J. Xiong, S. K. Kushwaha, T. Liang, J. W. Krizan, M. Hirschberger, W. Wang, R. J. Cava, N. P. Ong, Evidence for the chiral anomaly in the Dirac semimetal $Na_3Bi$. *Science* **350**, 413- 416(2015).

[27] X. Huang, L. Zhao, Y. Long, P. Wang, D. Chen, Z. Yang, H. Liang, M. Xue, H. Weng, Z. Fang, X. Dai, G. F. Chen, Observation of the Chiral-Anomaly-Induced Negative Magnetoresistance in 3D Weyl Semimetal TaAs. *Phys. Rev. X* **5**, 031023 (2015).

[28] G. Kresse and J. Furthmüller, Efficiency of Ab-Initio Total Energy Calculations for Metals and Semiconductors Using a Plane-Wave Basis Set, *Comput. Mater. Sci*. **6**, 15 (1996).

[29] G. Kresse and J. Furthmüller, Efficient Iterative Schemes for Ab Initio Total-Energy Calculations Using a Plane-Wave Basis Set, *Phys. Rev. B* **54**, 11169 (1996).

[30] J. P. Perdew, K. Burke, and M. Ernzerhof, Generalized Gradient Approximation Made Simple, *Phys. Rev. Lett.* **77**, 3865 (1996).

[31] R. Yu, X.L. Qi, A. Bernevig, Z. Fang, and X. Dai, Equivalent Expression of $Z_2$ Topological Invariant for Band Insulators Using the Non-Abelian Berry Connection, *Phys. Rev. B* **84**, 075119 (2011).

[32] Q. S. Wu and S.N. Zhang, Wannier Tools: An open-source software package for novel topological materials. http://github.com/quanshengwu/wannier_tools.

# Acknowledgements

**Funding:** This work was supported by the Ministry of Science and Technology of China (2016YFA0300600, 2016YFA0401000, 2016YFA0302400, 2015CB921300, and 2013CB921700, 2016YFA0300300), the National Natural Science Foundation of China (11622435, 11474340, 11422428, 11674369, 11234014, and 11674371), the Chinese Academy of Sciences (XDB07000000). Y.-B. H. acknowledges funding from the CAS Pioneer Hundred Talents Program (type C).

**Author contributions:** H.D. and T.Q. conceived the ARPES experiments; J.Z.M. and T.Q. performed ARPES measurements with the assistance of B.Q.L., L.Y.K., X.G., L.Y.R, and Y.B.H.; Y.F.X. and H.M.W. performed *ab initio* calculations; J.B.H., D.C., W.L.Z., and G.F.C. synthesized the single crystals; S.Z., D.C., C.Y.X., and G.F.C. performed quantum oscillation measurements; J.Z.M., T.Q., and H.D. analyzed the experimental data; J.Z.M., Y.F.X., T.Q., and J.B.H. plotted the figures; X.D. discussed the experimental and calculated data; Y.S., J.Z.M., Y.L.W., and H.J.G. performed STM experiments. T.Q., C.F., H.M.W., H.D., J.Z.M., and P.R. wrote the manuscript.

**Competing interests:** The authors declare that they have no competing interests.

**Data and materials availability:** All data needed to evaluate the conclusions in the paper are present in the paper and/or the Supplementary Materials. Additional data related to this paper may be requested from the authors.

# List of Supplementary Materials

Materials and Methods
Supplementary Text
Figs. S1 and S2
References (28-32)

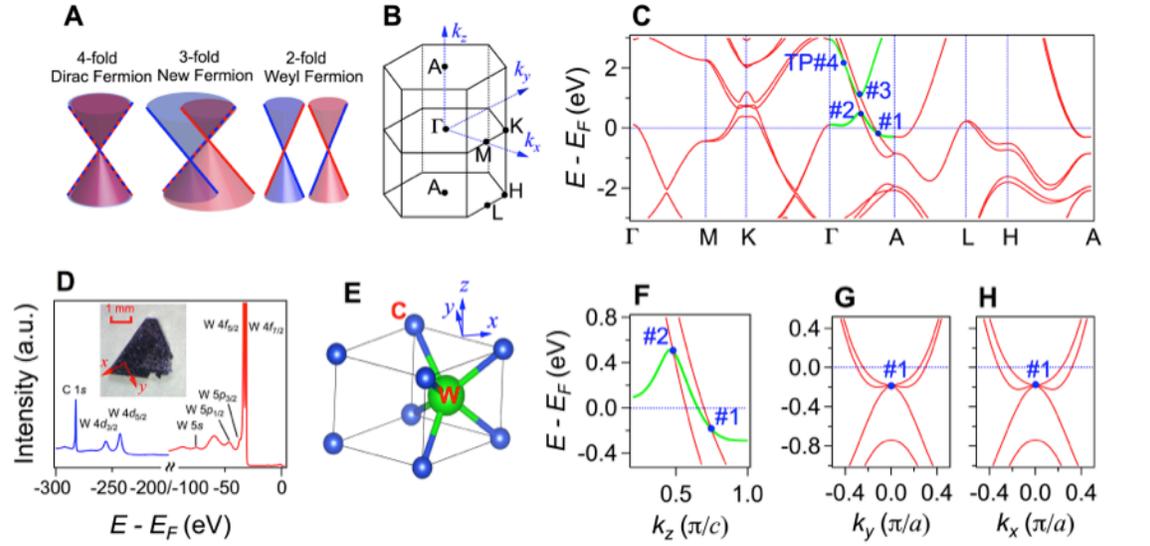

**Fig. 1. Electronic structure of TPs in WC.** (A) Schematic plots of band structures of 4-fold Dirac fermion, 3-fold new fermion, and 2-fold Weyl fermion. (B) Bulk Brillouin zone (BZ) with high-symmetry points. (C) Calculated bulk band structure along high-symmetry lines. The blue points indicate the four TPs and the green curves represent the doubly-degenerate bands along Γ-A. (D) Core level photoemission spectrum showing the characteristic peaks of W and C elements. The inset shows a typical triangular single crystal. (E) Crystal structure of WC in one unit cell. (F) to (H) Calculated bands along $k_z$, $k_y$, and $k_x$ through TP #1, respectively.

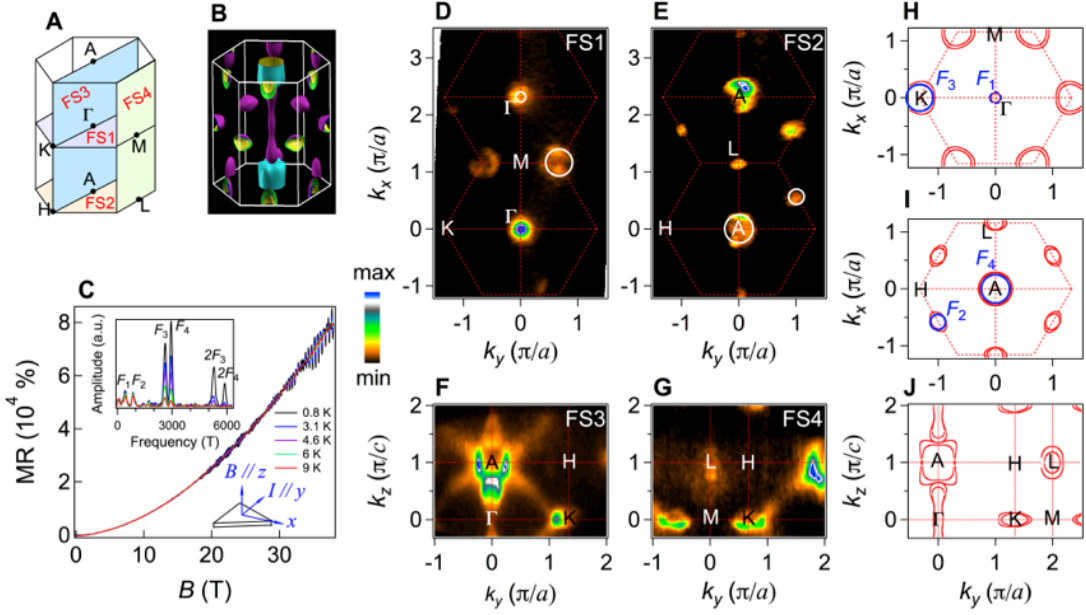

**Fig. 2. FSs in the (001) and (100) planes of WC.** (A) Bulk BZ with the four planes indicating the locations of the measured FSs in (D) to (G). (B) Calculated bulk FSs in the 3D bulk BZ. (C) Magnetoresistance (MR) as a function of magnetic field $B$ at different temperatures, in which the ripples show quantum oscillations. The electric current $I$ and $B$ are parallel to the [010] and [001] directions, respectively. The inset plots fast Fourier transforms of the MR data, showing four principal frequencies, $F_1$ to $F_4$. (D) to (G) Photoemission intensity plots at $E_F$ in the four high-symmetry planes indicated in (A). (H) to (J) Calculated bulk FSs in the four high-symmetry planes. The white circles in (D) and (E), and blue circles in (H) and (I), sketch the area of FSs extracted from the quantum oscillation data in (C).

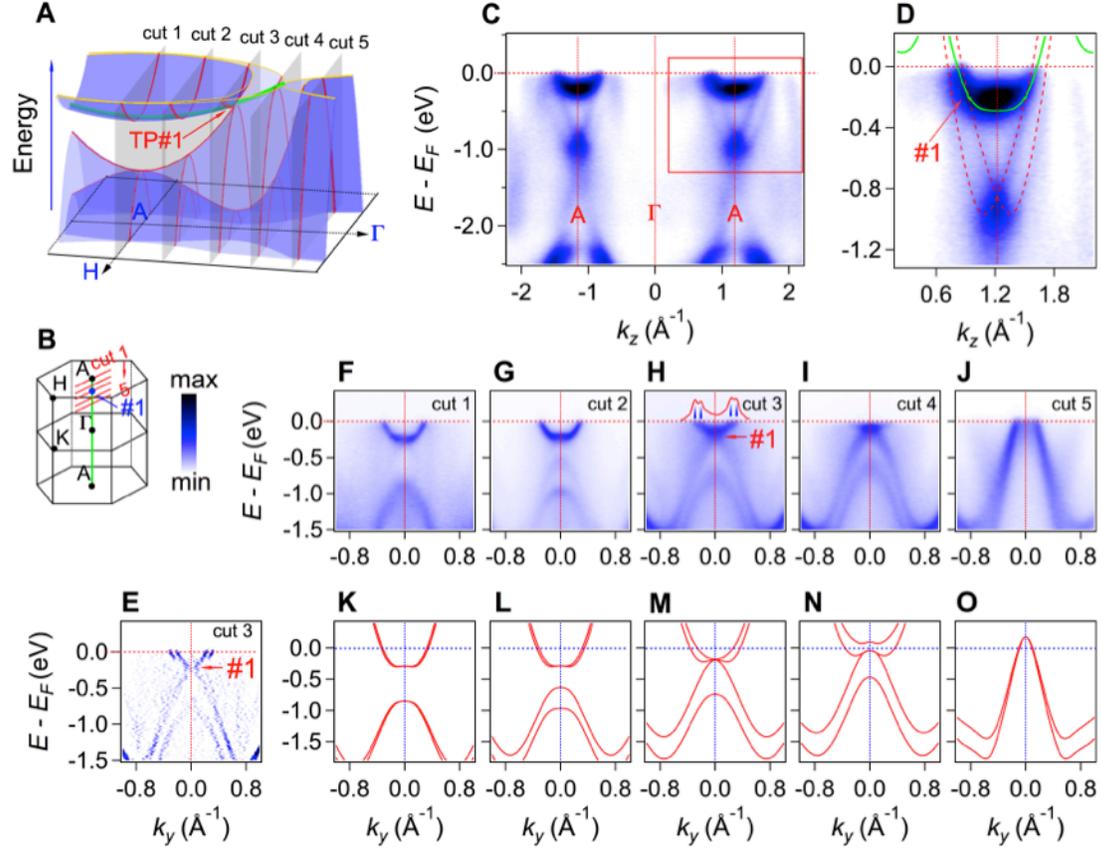

**Fig. 3. Band dispersions near TP #1.** (A) 3D schematic plot of the electronic band structure near TP #1 in the $k_x = 0$ plane. The gray planes indicate the momentum locations of cuts #1 to #5 in (F) to (J). The red and green curves represent the non-degenerate and doubly-degenerate bands, respectively. The yellow curves indicate the FSs. (B) 3D bulk BZ with red and green lines indicating the locations of cuts 1 to 5 in (F) to (J) and the cut in (C), respectively. (C) Photoemission intensity plot of band dispersions along A-Γ-A. (D) Zoom-in of the region in the red box in (B) with the calculated bands on top of the data. (E) Curvature intensity plot of (H) showing two electron bands. (F) to (J) Photoemission intensity plots of the band dispersions along cuts 1 to 5 indicated in (A), respectively. The momentum distribution curve at $E_F$ in (G) shows double peaks, which is consistent with two electron bands in (E). (K) to (O) Calculated bulk bands along cuts 1 to 5, respectively.

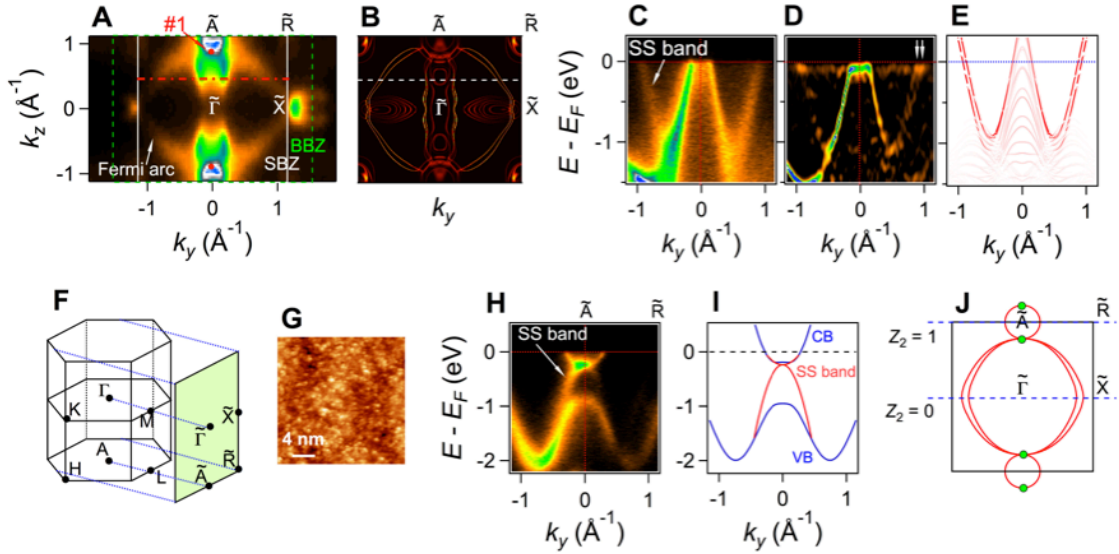

**Fig. 4. SSs on the (100) surface of WC.** (A) Photoemission intensity plot at -200 meV recorded with a photon energy of 555 eV, which measures the bulk electronic structure at $k_x = 0$. The red points represent TP #1 at $k_z$ and $-k_z$. The white and green rectangles indicate the (100) surface BZ (SBZ) and bulk BZ (BBZ), respectively. (B) Calculated SS FSs at the energy of TP #1 for a 15-unit-cell-thickness (100) slab with the carbon-terminated layer. (C) ARPES intensity plot showing band dispersions along the horizontal red line in (A). (D) Curvature intensity plot of the data in (C). (E) Calculated SS bands along the horizontal white line in (B). (F) Bulk and (100) surface BZ. (G) STM constant current topographic image of the *in-situ* cleaved (100) surface under ultra high vacuum, scanning parameters: $V = -1.0$ V, $I = 0.4$ nA. (H) ARPES intensity plot showing band dispersions along $\tilde{A} - \tilde{R}$. (I) Schematic band dispersions along $\tilde{A} - \tilde{R}$ showing a Dirac SS band connecting the bulk valence band (VB) and conduction band (CB). (J) Schematic plot of surface Fermi arcs emanating from the surface projection of TP #1 (green dots), which is consistent with the $Z_2$ numbers at the $k_z = 0$ and $\pi$ planes.